\documentclass[prl,superscriptaddress,amsmath,amssymb,floatfix,twocolumn]{revtex4-1}
\usepackage{graphicx,color}
\usepackage{amsmath,amsfonts,amssymb}
\usepackage{float}
\usepackage[colorlinks,linkcolor=blue,anchorcolor=blue,urlcolor=blue,citecolor=blue]{hyperref}

\newcommand{\edit}[1] {\textcolor{black}{#1}}  

\begin{document}

\title{\edit{Majorana-like} zero modes in Kekul\'e distorted sonic lattices}

\author{Penglin Gao}
\affiliation{Department of Physics, Universidad Carlos III de Madrid, ES-28916 Legan\`es, Madrid, Spain}
\author{Daniel Torrent}
\affiliation{GROC, UJI, Institut de Noves Tecnologies de la Imatge (INIT), Universitat Jaume I, 12071, Castell\'o, Spain}
\author{Francisco Cervera}
\affiliation{Wave Phenomena Group, Department of Electronic Engineering, Universitat Polit\`ecnica de Val\`encia, Camino de vera s.n. (Building 7F), ES-46022 Valencia, Spain}
\author{Pablo San-Jose}
\email[Corresponding author.\\]{pablo.sanjose@csic.es}
\affiliation{Instituto de Ciencia de Materiales de Madrid (ICMM-CSIC), Sor Juana In\'es de la Cruz 3, 28049 Madrid, Spain. Research Platform for Quantum Technologies (CSIC).}
\author{Jos\'e S\'anchez-Dehesa}
\affiliation{Wave Phenomena Group, Department of Electronic Engineering, Universitat Polit\`ecnica de Val\`encia, Camino de vera s.n. (Building 7F), ES-46022 Valencia, Spain}
\author{Johan Christensen}
\email[Corresponding author.\\]{johan.christensen@uc3m.es}
\affiliation{Department of Physics, Universidad Carlos III de Madrid, ES-28916 Legan\`es, Madrid, Spain}
\date{\today}

\begin{abstract}
Topological phases have recently been realised in bosonic systems. The associated boundary modes between regions of distinct topology have been used to demonstrate robust waveguiding, protected from defects by the topology of the surrounding bulk. A related type of topologically protected state that is not propagating but is bound to a defect has not been demonstrated to date in a bosonic setting. 
Here we demonstrate numerically and experimentally that an acoustic mode can be topologically bound to a vortex fabricated in a two-dimensional, Kekul\'e-distorted triangular acoustic lattice. Such lattice realises an acoustic analogue of the Jackiw-Rossi mechanism that topologically binds a bound state in a p-wave superconductor vortex. The acoustic bound state is thus a bosonic \edit{\emph{analogue}} of Majorana bound state, \edit{where the two valleys replace particle and hole components}. We numerically show that it is topologically protected against arbitrary symmetry-preserving local perturbations, and remains pinned to the Dirac frequency of the unperturbed lattice regardless of parameter variations. We demonstrate our prediction experimentally by 3D printing the vortex pattern in a plastic matrix and measuring the spectrum of the acoustic response of the device. Despite viscothermal losses, the measured topological resonance remains robust, with its frequency closely matching our simulations.
\end{abstract}

\maketitle

The possibility of engineering a special kind of electronic states of topological origin in condensed matter systems has spurred a revolution in the field \cite{Hasan:RMP10,Elliott:RMP15,Aguado:RNC17}. Topological states are robust against microscopic perturbations. This basic property is known as topological protection, and is at the root of their potential for future applications in quantum technologies \cite{Acin:NJP18}. Topological states are universally bound to topological defects in gapped electronic phases. Such defects can be boundaries between regions of distinct band topology, as in the Quantum Hall or Quantum Spin Hall effect in two-dimensions, which yields extended (propagating) topological edge modes. Other topological defects are zero-dimensional, which yield topological bound states. A paradigmatic example is the case of Majorana states bound to vortices in spinless p-wave superconductors \cite{Alicea:RPP12}. Majorana bound states are very exotic from many points of view, and are often presented as the solution to scalable quantum computation \cite{Nayak:RMP08}. The topological protection of these new electronic states usually depends on specific symmetries being preserved in the system. In the case of Majorana bound states, this requires the mean field particle-hole symmetry intrinsic to superconducting phases.

\begin{figure}
	\centering
	\includegraphics[scale=1]{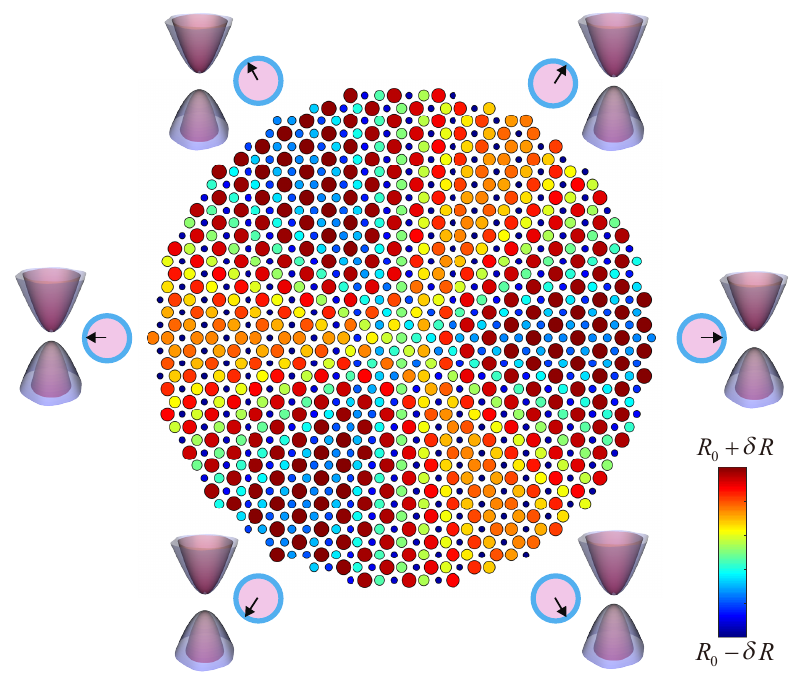}
	\caption{An acoustic analogue of a topological superconductor vortex made of rigid cylinders of varying radii, arranged in a finite triangular sonic lattice. The color of the cylinders illustrates the radius distortion $R_0\pm\delta R$, as given by Eq. (\ref{R}). The Kekul\'e phase $\phi$, together with the local intervalley gap as one moving around the vortex, are depicted around the perimeter.}
	\label{fig1}
\end{figure}

Sonic lattices have long been studied as a way to engineer the band structure of sound in a number of ways \cite{PhysRevLett.71.2022,Meseguer}. Recently, this has been extended to topological band structures, which has lead to the demonstration of topologically protected acoustic edge/interfacial channels \cite{yang2015topological,he2016acoustic,NatPhys.13.369,deng2017observation,PhysRevLett.120.246601,PhysRevApplied.9.034032,TopoSound}. The challenge of creating topologically bound acoustic modes, analogous to Majorana bound states, remains open. In this Letter we present a recipe to achieve this goal by implementing the analogue of a Jackiw-Rossi vortex \cite{Jackiw:NPB81,vortex1,vortex2} in a two-dimensional triangular sonic lattice made of rigid cylinders at a distance $d$. Such a lattice, of lattice constant $a = \sqrt{3}d$, exhibits two Dirac cones, or `valleys', at wavevectors $\textbf{K}_{\pm}=[\pm 4\pi/3d, 0]$, and at a Dirac frequency $\Omega_D$, which has to be computed numerically \cite{Torrent:PRL12}. The two Dirac cones are in principle independent and gapless, but by adding a Kekul\'e distortion \cite{Kekule:ACP66} to the cylinder radii they become coupled, and a gap $\Delta$ opens at $\Omega_D$. The effective low energy Hamiltonian of sound then takes a form analogous to the electronic Dirac equation with an intervalley gap $\Delta$
\begin{equation}
H = \int d^2r 
\sum_{s=\pm}\psi^\dagger_s\left(-isv\vec{\sigma}\cdot\vec{\partial}_r\right)\psi_s + \Delta\psi^\dagger_+\psi_{-} + \Delta^*\psi^\dagger_-\psi_{+}
\end{equation}
where $\hbar=1$, $v$ is a velocity, $\psi_\pm$ are the sound amplitudes on each valley, $s=\pm$ denotes the two valleys, and $\vec{\sigma}$ are Pauli matrices for the Dirac pseudospin (see Supplementary Material). Jackiw and Rossi demonstrated that making the complex intervalley coupling $\Delta$ position dependent, and by implementing a vortex in its phase around a given point, a topological bound state develops at the vortex core. In the fermionic case, $\psi_{\pm}$ are Dirac fields related by charge conjugation, $\psi_{-}=-i\psi_{+}^\dagger\sigma_2$, and as a result the bound state is a Majorana anyon \cite{Read:PRB00,Fu:PRL08,Nishida:PRB10}. \edit{While this relation does not hold in our case}, we show below that a Kekul\'e vortex in the sonic lattice also binds a topological state by the \edit{same} Jackiw-Rossi mechanism. 

\edit{A similar idea was put forward in graphene by Hou \emph{et al.} \cite{Hou:PRL07}, although engineering Kekul\'e distortions in graphene remains an open challenge. It is important to emphasize that in our acoustic implementation, like in the graphene proposal, the charge-conjugation relation does not hold, so this bound state is not an actual Majorana (i.e. an equal-weight superposition of particles and holes). It is, however, a `fractionalized' state \cite{Hou:PRL07} of topological origin, the closest possible acoustic analogue of Majorana bound state, with the crucial Majorana self-conjugation property replaced by an exactly vanishing valley polarization at any point $\textbf{r}$, i.e. an equal weight for $|\psi_+(\textbf{r})|$ and $|\psi_-(\textbf{r})|$ components in the wavefunction.}

Figure \ref{fig1} shows a sketch of our sonic lattice, a finite cluster of rigid cylinders forming a triangular lattice. On this lattice we implement intervalley coupling through a Kekul\'e distortion that consists of a position-dependent variation of the cylinder radius $R(\textbf{r})$ 
\begin{equation}
R(\textbf{r})=R_0+\delta R(r) \cos\left[\textbf{K}\cdot\textbf{r} + \phi(\textbf{r})\right]
\label{R}
\end{equation} 
where $\textbf{K}=\textbf{K}_{+}-\textbf{K}_{-}$, $R_0$ stands for the undisturbed radius, $\delta R(r)$ is the radius variation and $\phi(\textbf{r})$ is a phase.

\begin{figure}
	\centering
	\includegraphics[scale=1]{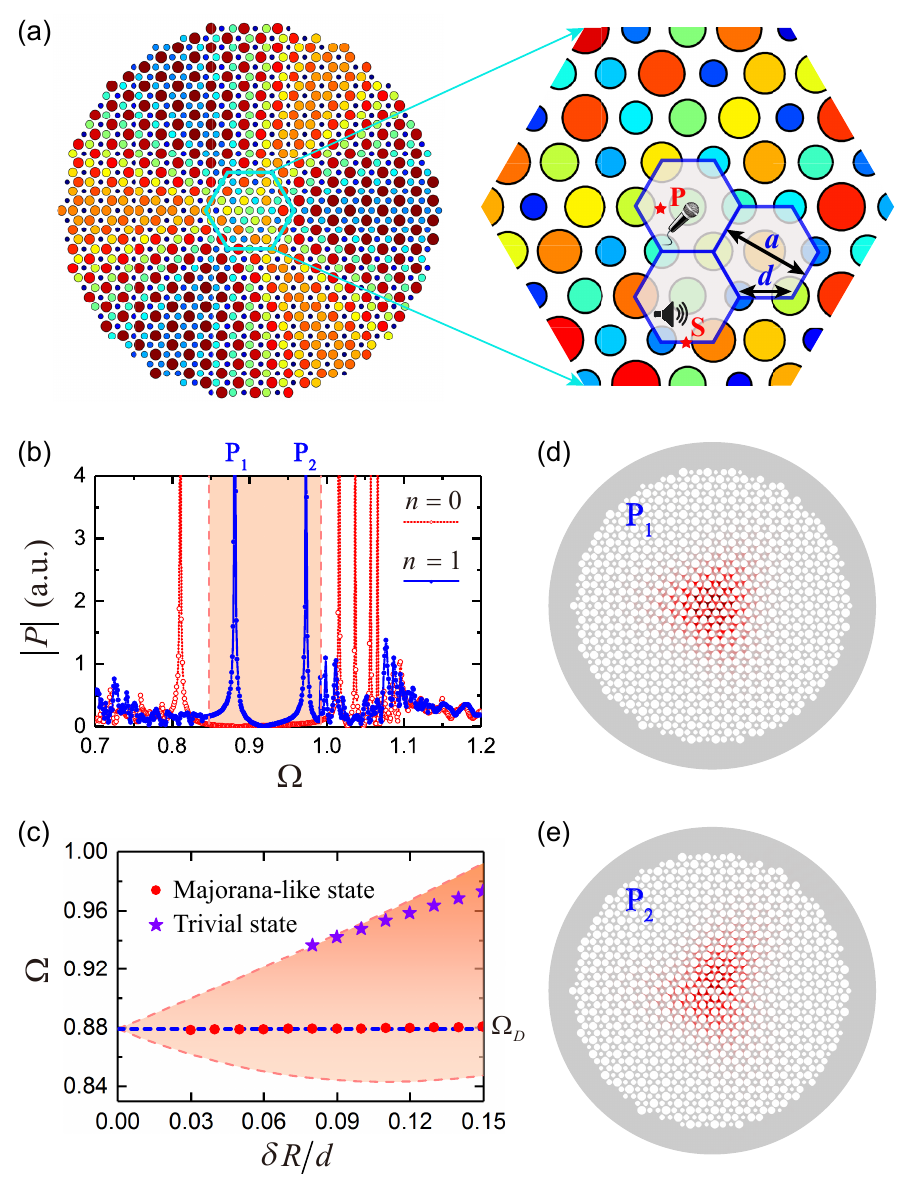}
	\caption{(a) Schematic of the cluster with a magnification of its core showing the locations of sound excitation (S) and probing (P). (b) Calculated pressure $|P|$ spectra for a topological trivial ($n=0$) and nontrivial ($n=1$) cluster, with  $\delta R/d=0.15$. The shaded background represents the complete bandgap. Frequency $\omega$ is normalized as $\Omega=\omega a/2\pi c$, with $c$ being the speed of sound and $a$ the lattice period. (c) Evolution of the localized states as a function of the distortion $\delta R/d$. The horizontally dashed line indicates the Dirac frequency, $\Omega_D=0.88$. (d) and (e) map the pressure amplitudes $|P|$ in real space for the two peaks $P_1$ and $P_2$ of panel (b).}
	\label{fig2}
\end{figure}

\begin{figure*}
	\centering
	\includegraphics[width=1.0\textwidth]{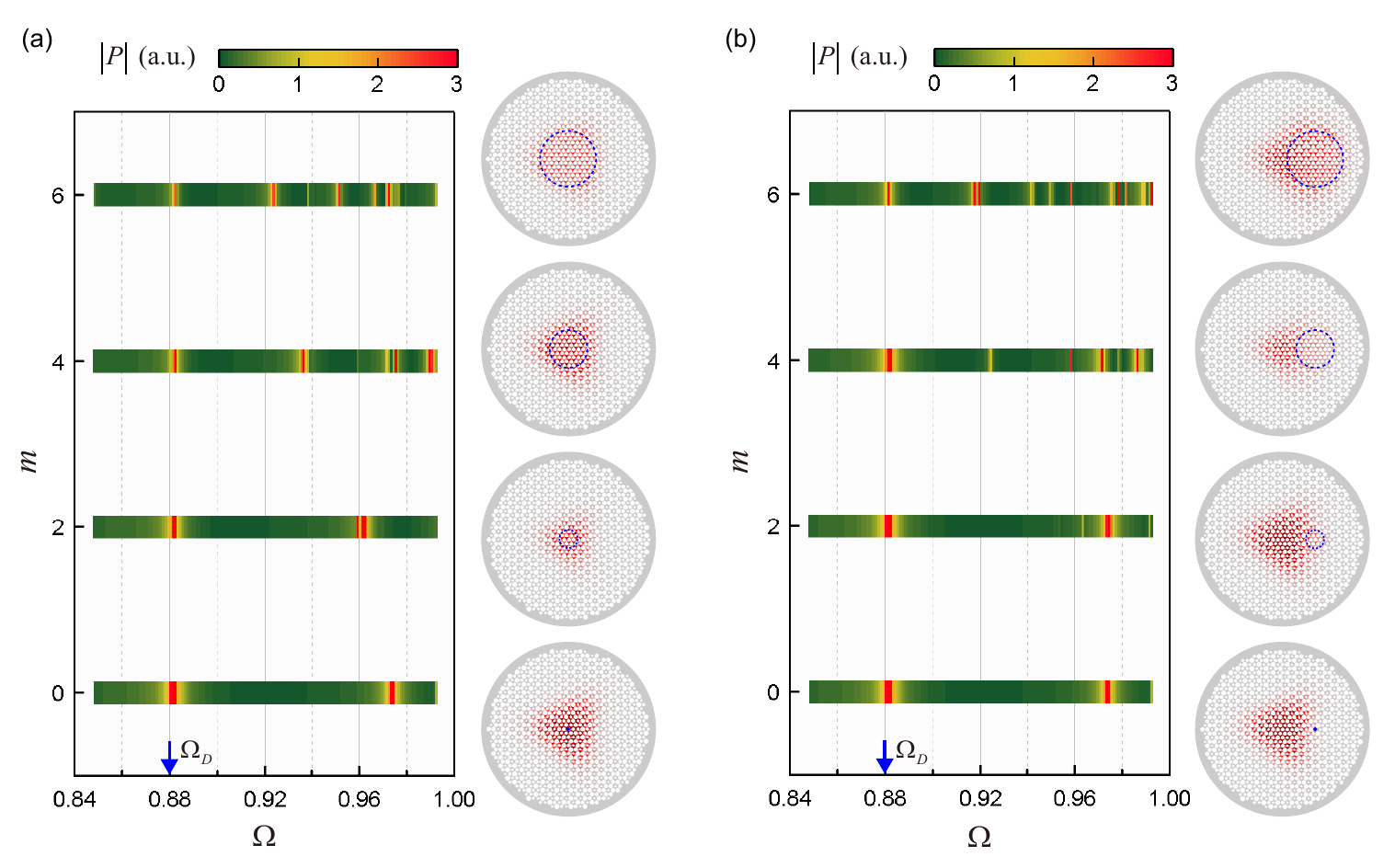}
	\caption{Response spectra $|P(\Omega)|$ within the bandgap of the topological $n=1$ vortex cluster ($\delta R/d=0.15$) with defect areas introduced (a) at the vortex core and (b) slightly away from it. The defects perturb the vortex by making $\delta R=0$ inside the dashed circles.  The different data sets to the left of each panel correspond to an increasing normalised radius $R_D/d = m$ of the circular defect area, while the spatial map to the right shows in red the mode profile of state $P_1$, which is found to remain pinned at the Dirac frequency $\Omega_D=0.88$ for all defects.}
	\label{fig3}
\end{figure*}

\begin{figure}
	\centering
	\includegraphics[scale=1]{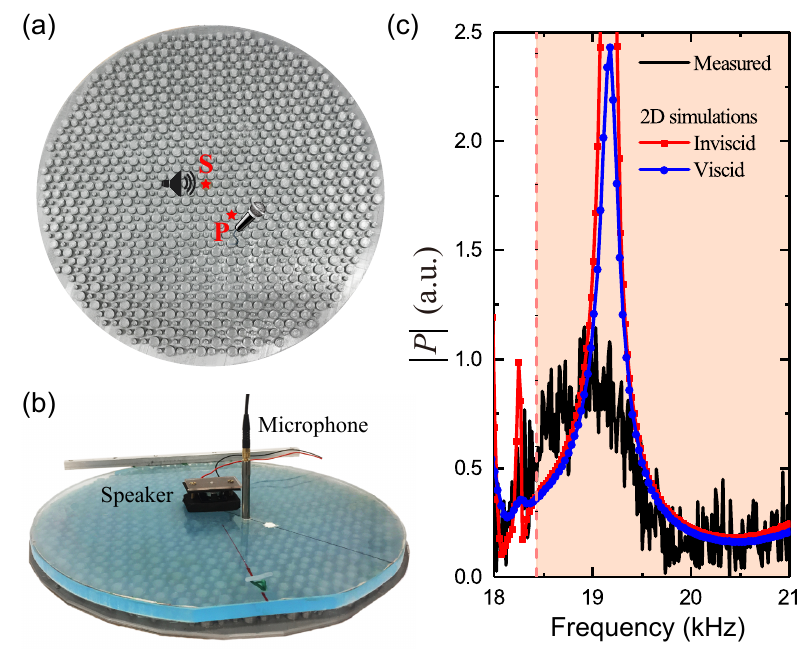}
	\caption{(a) Photograph of the 3D printed Kekul\'e distorted sonic lattice, with parameters $n=1$, $d=9\,\text{mm}$, $R_0=0.35d$, $\delta R=0.15d$, $\xi=2a$, and $a=\sqrt{3}d$. (b) A rigid lid was mounted above the cluster into which holes were drilled to fasten the speaker and the microphone. (c) The spectrally measured pressure at point $P$ (black) is compared to both inviscid (red) and viscid (blue) numerical computations in near proximity to the edge of the bandgap.}
	\label{fig4}
\end{figure}

In the pristine lattice with uniform $R(\textbf{r})=R_0$ the two Dirac cones are gapless. If we consider a unit cell with three cylinders, matching the Kekul\'e periodicity, the Brillouin zone folds the two Dirac cones onto the $\Gamma$ point. A Kekul\'e distortion with spatially uniform $\delta R$ and $\phi$ produces an intervalley coupling $\Delta\sim e^{i\phi}\delta R$ that lifts their degeneracy and opens a bandgap around the Dirac point (see the Supplementary Material). The Jackiw-Rossi vortex modeled in Fig. \ref{fig1}, in contrast, is a non-periodic configuration realised by position-dependent perturbation $\delta R(r)=\delta R \tanh (r/\xi)$ for a certain vortex radius $\xi$, and $\phi(\textbf{r})=\phi(x,y)=n\arctan(y/x)$, where $n$ is called the winding number, and is $n = 1$ in our case. Crucially, the intervalley phase $e^{i\phi(\textbf{r})}$ has winding number $n\neq 0$ around the vortex core, which makes the vortex a topologically non-trivial defect.


In what follows, we show, by a combination of simulations and experiments, that the artificial cluster with the $n=1$ Kekul\'e vortex described above supports an acoustic Majorana-like state pinned at the Dirac frequency, and realises the Jackiw-Rossi binding mechanism in an acoustic setting. The $\tanh$ radial profile of the vortex makes $\delta R(0)=0$ at the core, ensuring continuity. The core is thus locally gapless, while its contour remains locally gapped at the Dirac frequency $\Omega_D$, as depicted in the perimeter of Fig. \ref{fig1}. A topological state becomes bound to this gapless core, but due to the winding of the surrounding intervalley coupling its frequency is not shifted by the confinement. It thus remains exactly at $\Omega_D$ regardless of the value of the confinement radius $\xi$, as long as the cluster radius is much greater than $\xi$. This property is unique to topological confinement, and is the reason that Majorana bound states have zero energy in superconductors. 
A second essential property of topologically bound states is topological protection. Similar to the case of Majoranas in vortices of p-wave superconductors, we expect that our bound state will remain pinned at the Dirac point under any applied perturbation, as long as it preserves the particle-hole symmetry associated to its topological class. In other words, the acoustic counterpart to the particle-hole symmetry, when preserved, should keep the topological bound state entirely robust and fixed at $\Omega_D$ regardless of any symmetry-preserving defects in the cluster.

To demonstrate these two effects we simulate, using the COMSOL Multiphysics software, the excitation of different clusters by a monopole source (point $S$) close to its center, see Fig. \ref{fig2}(a). We probe the response of the system in the vicinity of the vortex core, located at $P$. We compare the spectral response of a topologically trivial gapped cluster without a vortex ($n=0$) and the nontrivial ($n=1$) cluster of Fig. \ref{fig1}. The specific parameters are $R_0=0.35d$, $\delta R=0.15d$ and $\xi=2a$. As expected, the $n=0$ trivial cluster exhibits a bandgap [orange region in Fig. \ref{fig2}(b)].  The nontrivial system shows a similar gap, but in addition binds two strongly localized states, labeled $P_1$ and $P_2$, within the vortex at frequencies inside the bandgap. Their spatial profile is shown in Figs. \ref{fig2}(d) and \ref{fig2}(e). We note that while both states are strongly bound, the confinement of $P_2$ state is slightly less than the $P_1$ state, and shows slower decay along certain directions.

At this point, it is not clear which one of these states, if any, is of topological origin and which one is an ordinary defect-like state. Interestingly though, the $P_1$ peak appears exactly at the Dirac frequency $\Omega_D=0.88$. In order to elucidate their origin we study their normalised frequency $\Omega$ as a function of the normalised Kekul\'e distortion $\delta R/d$, see Fig. \ref{fig2}(c). As we gradually increase the distortion, the state $P_1$ (solid dots) remains always inside the gap, and is found to emerge directly from the Dirac point. This is consistent with a $P_1$ state of topological origin.
(Note that it is not possible to reduce the modulation to $\delta R/d\lesssim 0.03$ without increasing the size of our simulation cluster, as the gap becomes so small that the confined state leaks out of the sample.) 
In contrast, the second peak $P_2$ emerges from bulk band only for a sufficiently strong distortion $\delta R/d\gtrsim 0.08$. It is thus a  trivial defect state, analogous to a Matricon-Caroli excitation \cite{Caroli:PL64} in a p-wave superconductor vortex. 

We now demonstrate that the $P_1$ state also enjoys the second property of topologically bound states, topological protection against local defects. In contrast, the trivial $P_2$ state is frail.
To this end we create local defects in the vortex cluster, in the form of gapless circular regions of arbitrary size and position close to the core, wherein $\delta R=0$. The defects disrupt the vortex locally, while preserving the requisite particle-hole symmetry, i.e. without shifting the Dirac frequency $\Omega_D$. The frequency of a true topological vortex state should thus remain completely insensitive to them.

The simulated defect regions are marked with blue dashed lines in Fig. \ref{fig3}. The defect radii $R_D = md$ are labeled by an integer $m$. Their centers are chosen either at the core [panel (a)] or displaced [panel (b)]. The corresponding spectral responses are shown to the left of each defect, and reveal that regardless of defect parameters, the $P_1$ state remains strictly pinned to the Dirac frequency $\Omega_D$. This is true even though its spatial profile (shown in red in the spatial sketches) exhibits strong variations due to the defects. Other peaks of trivial origin and varying frequency, similar to $P_2$, are found to enter the gap, particularly for the larger defects. 
The remarkable resilience of $P_1$ against arbitrary defects once more confirms that it is fundamentally different from all other states, and that it is produced by topological confinement in the particle-hole symmetric vortex. Its general behaviour, as shown in Figs. \ref{fig2} and \ref{fig3}, is analogous to that of a Majorana bound state in a p-wave superconductor vortex.

Finally, we have fabricated a physical realisation of our Kekul\'e vortex by 3D printing the cluster pattern of Fig. \ref{fig1} in a plastic matrix, and measuring its acoustic response. A photograph of the sample is shown in Fig. \ref{fig4}(a). The sample has a diameter of 325 mm and a cylinder height of 5 mm. On top of the plastic cylinders that can be considered acoustically rigid we attached a transparent lid, into which holes were drilled to hold a microphone (probe, $P$) and the speaker (source, $S$), see Fig. \ref{fig4}(b). We used a National Instruments multifunction DAQ card PXIe-6259 to generate the excitation signal as well as to measure and analyse the acquired response signal. A SAMSOM SA200 Power Amplifier was used to amplify the excitation chirp signal emitted through a AIYIMA 23 mm diameter speaker.
The receiving signal was detected through a Bruel \& Kjaer 4954 $1/4$ inch microphone,  located 35 mm away from the speaker. The data were collected and digitally processed in a computer. In order to avoid unwanted reflections at the open rim of the 2D cavity (as formed between the lid and the cluster) we surrounded the sample with absorbing foam. The narrow channels between neighbor cylinders and also the gap between the lid and the free upper cylinder-surface, constitute an unavoidable source for high viscothermal losses in the crystal lattice, against which a topological state is unfortunately not expected to be protected. 

Despite the viscous losses, we have successfully detected the \edit{topological} $P_1$ peak in our sample, see black curve in Fig. \ref{fig4}(c). The peak is considerably broadened due to viscous losses, however. The numerical simulations presented until now were performed for inviscid fluids. We have repeated them including viscous and thermal loss mechanisms in the 2D linearized Navier-Stokes equations. The measured and calculated spectra are compared in Fig. \ref{fig4}(c). The experimentally observed bound state peaks at around 19.03 kHz in comparison to the predicted one, located at 19.18 kHz. Due to the finite height of the cylinders an additional viscous boundary layer is formed on top of them, which is expected to be responsible for additional broadening, not captured by the 2D simulations. Temperature variations compared to the simulated results may also account for the slight frequency shift seen in Fig. \ref{fig4}(c). 

In conclusion, we have demonstrated the topological binding of a \edit{fractionalized acoustic mode}, a sonic analogue of a Majorana state in a Kekul\'e distorted triangular lattice with a vortex in the Kekul\'e phase. By deliberately introducing symmetry-preserving defects in the form of large areas of substitutional cylinders, we show that the state is topologically protected, and remains pinned to the Dirac frequency even in the presence of strong defects. Finally, we have created a physical realisation of the acoustic Kekul\'e vortex by 3D printing, and have detected the Majorana-like topological acoustic bound state at a frequency closely matching our simulations. This constitutes the first direct evidence of a \edit{topological bound acoustic state}.  Beyond its fundamental significance, it is expected to inspire new routes towards the engineering of topologically protected sonic states, their manipulation and transmission.

\begin{acknowledgments}
J. C. acknowledges the support from the European Research Council (ERC) through the Starting Grant No. 714577 PHONOMETA and from the MINECO
through a Ram\'on y Cajal grant (Grant No. RYC-2015-17156). J.S-D. acknowledges the support by the Ministerio de Econom\'ia y Competitividad of the Spanish Goverment and the European Union "Fondo Europeo de Desarrollo Regional (FEDER)" through project No. TEC2014-53088-C3-1-R. P.S-J. acknowledges support from MINECO/FEDER under grant No. FIS2015-65706-P.
\end{acknowledgments}

\bibliography{references}

\end{document}